\documentclass[final,5p,times,twocolumn,authoryear]{elsarticle}

\usepackage{amssymb}
\usepackage{amsmath}
\usepackage[authoryear]{natbib}
\usepackage{framed} 

\usepackage{multicol} 
\usepackage{multirow}
\usepackage{array}
\usepackage{booktabs}

\usepackage{nomencl} 
\usepackage{threeparttable}
\usepackage{hyperref}

\makenomenclature

\renewcommand*\nompreamble{\begin{multicols}{2}}

\renewcommand*\nompostamble{\end{multicols}}

\usepackage{pifont}
\newcommand{\xmark}{\ding{55}}  
\newcommand{\cmark}{\ding{51}}  

\usepackage{tabularx}
\usepackage{array}
\newcolumntype{L}[1]{>{\raggedright\arraybackslash}p{#1}}
\usepackage{microtype}
\usepackage{caption}
\usepackage{algpseudocode}
\algrenewcommand{\algorithmiccomment}[1]{\texttt{// #1}}
\usepackage[ruled,vlined]{algorithm2e}
\usepackage{graphicx}
\usepackage{amsthm}

\usepackage{etoolbox}
\renewcommand\nomgroup[1]{%
  \item[\bfseries
  \ifstrequal{#1}{A}{Abbreviations}{%
  \ifstrequal{#1}{S}{Symbols}{%
  }}%
]}

\journal{Control Engineering Practice}
\date{}
\begin{document}

\begin{frontmatter}



\title{Intelligent Self-tuning Active EMI Filtering for Electrified Automotive Power Systems Using Reinforcement Learning} 


\author[1]{Mahuizi Lu}

\affiliation[1]{organization={Department of Earth Science Engineering},
    addressline={Imperial College London}, 
    city={London},
    postcode={SW7 2AZ}, 
    country={United Kingdom}}

\author[2]{Kelin Jia}
\affiliation[2]{organization={E-Powertrain, Jaguar Land Rover},
    city={Abbey Road, Whitley, Coventry},
    postcode={CV3 4LF}, 
    country={United Kingdom}}

\author[2]{Rajib Goswami}

\author[3,4]{Yukun Hu\corref{cor1}}
\ead{yukun.hu@ucl.ac.uk}

\affiliation[3]{organization={Department of Civil, Environmental \& Geomatic Engineering},
    addressline={University College London}, 
    city={London},
    postcode={WC1E 6BT}, 
    country={United Kingdom}}

\affiliation[4]{organization={Dynamic Systems Lab, Factulty of Engineering Sciences},
    addressline={University College London}, 
    city={London},
    postcode={WC1E 6BT}, 
    country={United Kingdom}}

\cortext[cor1]{Corresponding authors}

\begin{abstract}
The rapid electrification and intelligence of modern transportation systems place stringent demands on the electromagnetic compatibility, reliability, and adaptability of automotive power electronics. In electric and autonomous vehicles, electromagnetic interference (EMI) generated by high-frequency switching power converters can compromise safety-critical functions, in-vehicle communications, and system efficiency under dynamic operating conditions. Conventional passive EMI filters, while robust, are often overdesigned and lack adaptability, leading to increased weight, volume, and energy losses. This paper proposes an intelligent self-tuning active EMI filtering approach for electrified automotive power systems based on reinforcement learning (RL). The EMI mitigation problem is formulated as a Markov decision process, enabling an RL agent to continuously adapt filter parameters in response to time-varying interference characteristics. To improve robustness and generalisation under complex and non-stationary conditions, a variational autoencoder is employed for compact state representation, while a noise-based exploration mechanism enhances learning efficiency and prevents suboptimal convergence. The proposed method is evaluated using experimentally measured EMI spectra from an automotive electric drive unit within a MATLAB/Simulink co-simulation framework. Results demonstrate consistent EMI attenuation improvements of 25–30 dB across a wide frequency range compared with conventional control strategies and passive filtering solutions. By reducing reliance on oversized passive components and enabling adaptive EMI suppression, the proposed framework supports lightweight, energy-efficient, and reliable power-electronic systems for intelligent and green transportation applications.
\end{abstract}






\end{frontmatter}
\begin{table*}[!t]   
\begin{framed}
\nomenclature[A]{EMI}{Ectromagnetic Interference}
\nomenclature[A]{PEFs}{Passive EMI Filters}
\nomenclature[A]{AEFs}{Active EMI Filters}
\nomenclature[A]{RL}{Reinforcement Learning}
\nomenclature[A]{MDP}{Markov Decision Process}
\nomenclature[A]{VAE}{Variational Autoencoder}
\nomenclature[A]{EQRL}{Enhanced Q-value-based RL}
\nomenclature[A]{GSA}{Gravitational Search Algorithm}
\nomenclature[A]{ACO}{Ant Colony Optimisation}
\nomenclature[A]{CM}{Common-Mode}
\nomenclature[A]{DM}{Differential-Mode}
\nomenclature[A]{MLP}{Multilayer Perceptron}
\nomenclature[A]{ReLU}{Rectified Linear Unit}
\nomenclature[A]{TD}{Temporal Difference}
\nomenclature[A]{EDU}{Electric Drive Unit}
\nomenclature[A]{FFT}{Fast Fourier Transform}
\nomenclature[A]{LV}{Low Voltage}
\nomenclature[A]{CT}{Current Transformer}
\nomenclature[A]{LISN}{Line Impedance Stabilisation Network}
\nomenclature[A]{DC}{Direct Current}
\nomenclature[A]{RMSE}{Root Mean Squared Error}
\nomenclature[A]{SARSA}{State Action Reward State Action}
\nomenclature[A]{Op-Amps}{Operational Amplifiers}
\nomenclature[A]{GA}{Genetic Algorithm}

\nomenclature[S]{$\textit{A}$}{set of all actions [-]}
\nomenclature[S]{$C$}{capacitor [$F$]}
\nomenclature[S]{$f$}{frequency [$Hz$]}
\nomenclature[S]{$\mathcal{D}$}{replay memory [-]}
\nomenclature[S]{$G$}{gain [-]}
\nomenclature[S]{$I$}{current [$A$]}
\nomenclature[S]{$L$}{inductor [$H$]}
\nomenclature[S]{${m}$}{first-order moment estimate of TD error [-]}
\nomenclature[S]{$\hat{m}$}{bias-corrected first-order moment estimate [-]}
\nomenclature[S]{$\mathcal{N}$}{normal distribution [-]}
\nomenclature[S]{$\textit{P}$}{probability [-]}
\nomenclature[S]{$q_\pi(s, a)$}{state-action value function [-]}
\nomenclature[S]{$Q$}{action-value function [-]}
\nomenclature[S]{$\textit{r}$}{reward [-]}
\nomenclature[S]{$R$}{resistor [$\Omega$]}
\nomenclature[S]{$s$}{state space [-]}
\nomenclature[S]{$\textit{S}$}{set of all state space [-]}
\nomenclature[S]{$V$}{voltage [V]}
\nomenclature[S]{$y$}{target value [-]}
\nomenclature[S]{$a$}{action space [-]}
\nomenclature[S]{$\gamma$}{discount factor [-]}
\nomenclature[S]{$\delta$}{temporal-difference (TD) error [-]}
\nomenclature[S]{$\varepsilon$}{small positive constant value [-]}
\nomenclature[S]{$\epsilon$}{greedy policy [-]}
\nomenclature[S]{$\eta$}{noise [-]}
\nomenclature[S]{$\tau_{\mathrm{EMI}}$}{target EMI threshold [dB$\mu$A]}
\nomenclature[S]{$\lambda$}{decay [-]}
\nomenclature[S]{$\sigma$}{standard deviation [-]}
\nomenclature[S]{${v}$}{second-order moment estimate of TD error [-]}
\nomenclature[S]{$\hat{v}$}{bias-corrected second-order moment estimate [-]}
\nomenclature[S]{$v_\pi(s)$}{state-value function [-]}
\nomenclature[S]{$\pi$}{policy [-]}
\nomenclature[S]{$t$}{time [$s$]}
\nomenclature[S]{$Z$}{impedance [$\Omega$]}
\nomenclature[S]{$\kappa$}{capacitance ratio [-]}
\nomenclature[S]{$\mu$}{means [-]}
\nomenclature[S]{$z$}{latent variables [-]}
\nomenclature[S]{$\theta$}{weight [-]}
\nomenclature[S]{$N$}{number of samples in the signal [-]}
\nomenclature[S]{$x$}{discrete signal value [-]}
\nomenclature[S]{$\omega$}{window function [-]}
\nomenclature[S]{$n$}{discrete time index [-]}

\printnomenclature
\end{framed}
\end{table*}



\section{Introduction}
The rapid transition toward electrified, connected, and intelligent transportation systems has fundamentally reshaped the design requirements of automotive power electronics. Electric vehicles and autonomous driving platforms increasingly rely on high-power-density converters, fast-switching powertrains, and sophisticated in-vehicle communication systems to support energy-efficient propulsion, advanced sensing, and real-time decision-making. These developments are central to the broader goals of green mobility, including improved energy efficiency, reduced vehicle mass, and enhanced operational reliability. However, one critical challenge arising from these technological developments is managing electromagnetic interference (EMI) \citep{Wang2023}. Autonomous and connected vehicles, especially those featuring electric powertrains, demand stringent EMI control to ensure safety and operational functionality. Switch-mode power supplies, integral to these advanced systems, inherently generate significant high-frequency EMI radiation due to frequent switching operations \citep{Mainali2010}. Additionally, varying operating conditions, including power fluctuations, temperature variations, mechanical vibrations, and load shifts, necessitate highly adaptable EMI mitigation solutions.

Traditionally, passive EMI filters (PEFs) have been the dominant solution for EMI suppression due to their simplicity, linearity, and inherent reliability \citep{Kotny2011, Vedde2021}. PEFs rely exclusively on passive components—such as resistors, capacitors, and inductors—whose fixed impedance characteristics attenuate unwanted frequency components. However, this fixed-parameter nature imposes fundamental limitations. PEFs cannot amplify signals, often introduce signal attenuation and power losses, and exhibit limited selectivity and adaptability across wide frequency ranges \citep{Jiraprasertwong2015, Stella2022}. Achieving stronger EMI attenuation typically requires additional passive components \citep{Akagi2004}, resulting in increased size, weight, and cost, which are increasingly incompatible with modern high-density power-electronic designs \citep{Baumann2021, Hamza2013}. In electrified vehicles, these limitations directly translate into increased vehicle mass, reduced packaging flexibility, and degraded energy efficiency.

Active EMI filters (AEFs) have emerged as a promising alternative capable of overcoming many of these limitations \citep{Rivas2003}. By integrating active elements such as operational amplifiers and transistors with passive components \citep{vanWyk2005}, AEFs offer controllable gain, high input impedance, and low output impedance, enabling enhanced EMI suppression performance \citep{Narayanasamy2019}. The use of smaller inductors in AEFs further reduces physical footprint, parasitic effects, and improves high-frequency behaviour, which is critical for effective EMI mitigation \citep{Chen2009}. Moreover, through active feedback mechanisms, AEFs can generate frequency-selective cancelling signals, allowing them to respond more flexibly to diverse and evolving interference patterns \citep{Bendicks2022, Bendicks2020, Chen2006}. Despite the se  advantages, the performance of AEFs remains highly sensitive to component tolerances, environmental disturbances, and operating-point variations, making reliable self-tuning a persistent challenge. Conventional analogue design approaches, constrained by nonlinearity and complex multi-parameter interactions, often exhibit discrepancies between simulation-based tuning and real-world performance \citep{Bao2024}.

Existing AEF tuning strategies typically rely on fixed-parameter designs \citep{Madhuri2024} or adaptive control schemes that require prior system knowledge and manual configuration \citep{electronics13193889, Du2023}. Metaheuristic optimisa,tion methods, e.g., Genetic Algorithm (GA) \citep{Chen2019}, gravitational search algorithm (GSA) \citep{Chauhan2023} and Ant Colony Optimisation (ACO) \citep{SAKTHIVEL2015153}, offer greater flexibility but are inherently offline, computationally intensive, and unsuitable for real-time adaptation under rapidly changing EMI conditions. Supervised machine learning approaches \citep{Chen2019}, while effective in specific scenarios, depend on labelled datasets and often struggle to generalise to unseen disturbances. The detailed comparison of the proposed technique with existing AEF tuning techniques is shown in Table \ref{tab:rotatedAEF}.

\begin{table*}[h]
\centering
\caption{Comparison of AEF tuning techniques}
\begin{tabular}{@{}L{3cm} L{1.75cm} L{1.75cm} L{2cm} L{2cm} L{2.25cm} L{2.25cm}@{}}
\toprule
\textbf{Technique} &
\textbf{Online adaptation} &
\textbf{Handles nonstationary EMI} &
\textbf{ Model-free} &
\textbf{Exploration efficiency} &
\textbf{High-dimensional state handling} &
\textbf{Generalisation ability} \\
\midrule
Fixed-Parameter & \xmark & \xmark & N/A & N/A & N/A & N/A \\
Adaptive Control & \cmark (limited) & \xmark & N/A & N/A & N/A & N/A \\
Metaheuristics & \xmark & \cmark (offline) & \cmark & \xmark & \xmark & Moderate \\
Supervised ML & \xmark & \xmark & \xmark & \xmark & \cmark & Moderate \\
Proposed Technique & \cmark & \cmark & \cmark & \cmark & \cmark & \cmark \\
\bottomrule
\end{tabular}
\label{tab:rotatedAEF}
\end{table*}

To address these limitations, this paper proposes a self-tuning active EMI filtering framework based on model-free reinforcement learning (RL), enabling continuous adaptation through direct interaction with the EMI environment. The EMI suppression problem is formulated as a Markov decision process (MDP), allowing an RL agent to learn optimal tuning policies without explicit system modelling. To enhance robustness and generalisation in high-dimensional EMI environments, a variational autoencoder (VAE) is employed for unsupervised feature extraction and compact state representation. In addition, a noise network is integrated into the learning architecture to promote structured exploration and prevent premature convergence to suboptimal solutions. From an intelligent transportation perspective, the proposed framework enables automotive power systems to autonomously adapt to real-world driving and operating conditions. The high levels of electrification, stringent electromagnetic compatibility requirements, and strongly time-varying operating conditions characteristic of modern automotive platforms make them a particularly demanding testbed for adaptive EMI mitigation.  

The main contributions of this paper are summarised as follows:
\begin{itemize}
    \item An informatics-driven self-tuning AEF framework is proposed, which formulates adaptive EMI suppression as a Markov decision process and enables continuous online reconfiguration through model-free reinforcement learning.
    \item A latent-state reinforcement learning architecture is developed by integrating a variational autoencoder for unsupervised feature extraction with a noise-based exploration mechanism, improving robustness, generalisation, and learning efficiency in high-dimensional, non-stationary EMI environments.
    \item A practical voltage-sensing, current-controlled AEF circuit architecture is designed and used as the physical environment for training and evaluation, incorporating damping and compensation networks to ensure wide-band stability.
    \item Comprehensive simulation-based validation using experimentally measured EMI data from an automotive electric drive unit demonstrates consistent EMI attenuation improvements over conventional control strategies and baseline learning methods.
\end{itemize}

The remainder of this paper is organised as follows. Section 2 describes methodology, including the self-tuning AEF architecture and the formulation of the reinforcement learning framework. Section 3 details the experimental setup for modelling and simulations, EMI data processing, and training procedure. Section 4 presents the evaluation results and comparative analysis, and discusses future research directions. Section 5 summarises the key conclusions.

\section{Methodology}
\begin{figure}[h]
\centering
\includegraphics[width=0.475\textwidth]{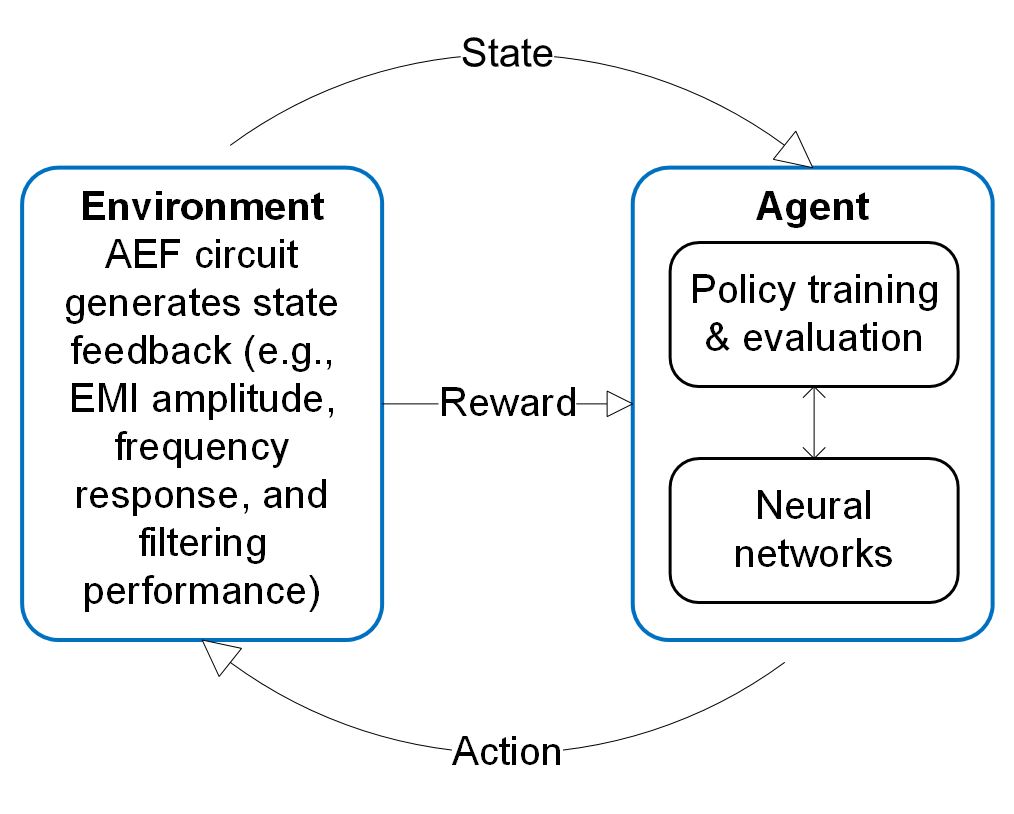}
\captionsetup{justification=centering} 
\caption{The schematic depicting the application of our proposed method workflow}
\label{method}
\end{figure}

This section provides a detailed overview of constructing a Self-Tuning AEF and its application within our proposed deep RL algorithm. We initially developed an AEF environment that provides real-time feedback on interference signals. This AEF system and its parameter configuration are then formulated as an MDP. The proposed method involves training an agent to develop an optimal offline-transferable strategy for dynamically adjusting the AEF parameters. The overall framework is illustrated in Figure \ref{method}.

\subsection{Self-tuning AEF System Design}

\label{sec:Self-tuning AEF System Design}
The AEF’s general structure is designed to minimise noise in motor drive systems. It actively suppresses conducted noise on the power line by generating compensating voltages or currents via a control amplifier. In this study, we adopt a voltage-sensing and current-control approach to optimise EMI suppression while maintaining a compact and efficient circuit structure. Voltage sensing eliminates the need for bulky current transformers and enables high sensitivity, while current control ensures sufficient bandwidth and current-handling capability to suppress both common-mode (CM) and differential-mode (DM) noise. The AEF is implemented using a feedback control strategy, which provides strong adaptability to dynamic EMI conditions. A Q-value-based RL algorithm is employed to tune the filter parameters in real time, including compensation impedance, feedback gain, and bandwidth. The overall method ensures robust noise suppression without compromising system stability.

In motor drive systems, EMI noise typically originates from both CM and DM sources, primarily due to high-speed switching and rapid current transitions. Effective filtering is necessary at the system interfaces to prevent conducted EMI from propagating through power cables into external circuitry. Our design incorporates a feedback-based AEF that dynamically adapts to these EMI conditions. However, even with feedback control, stability challenges can arise at both low and high frequencies due to parasitic effects and the inherent dynamics of operational amplifiers.

\begin{figure}[h]
\centering
\includegraphics[width=0.45\textwidth]{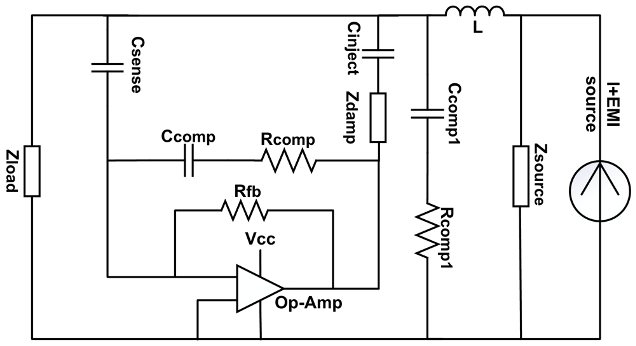}
\captionsetup{justification=centering} 
\caption{The circuit with damping and compensation components}
\label{circuit}
\end{figure}

To ensure robust and stable operation across a wide frequency range, we incorporate damping and compensation networks into the filter architecture. As shown in Figure \ref{circuit}, the feedback loop includes a multi-stage compensation network comprising $R_{\text{feedback}}$ for base feedback, $R_{\text{comp}}$ and $C_{\text{comp}}$ for low-frequency compensation, and $R_{\text{comp1}}$ and $C_{\text{comp1}}$ for high-frequency phase margin enhancement. Additionally, $Z_{\text{damp}}$ and $C_{\text{inject}}$ are included in the injection path to manage low-frequency resonance, while $C_{\text{sense}}$ provides voltage sensing feedback from the output node.

These compensation and damping techniques follow the methodology in \cite{Chu2016}.  
The low-frequency branch \((R_{\text{comp}}\,\parallel\,C_{\text{comp}})\) and the high-frequency branch \((R_{\text{comp1}}\,\parallel\,C_{\text{comp1}})\) shape the loop phase, while the series damper \(Z_{\text{damp}}\) suppresses low-frequency resonance.

The system’s switching frequency \(f_{\text{switching}}\) sets the usable control bandwidth, whereas the cut-off frequency \(f_{\text{cut-off}}\) separates actively compensated and naturally attenuated regions.  
Both parameters are iteratively tuned by the RL agent.

\paragraph*{Closed-loop transfer function}
For the voltage-sensing, current-controlled feedback AEF in Figure \ref{circuit},

\begin{equation}
    V_{o}(s)=V_{s}(s)\!\left[1-\frac{G(s)\,Z_{\text{feedback}}(s)}{Z_{s}+Z_{\text{feedback}}(s)}\right],
    \label{eq:VoVs}
\end{equation}

where  
\(V_s(s)\) is the noise-source voltage,  
\(V_o(s)\) is the output voltage across \(Z_{\text{load}}\),  
\(Z_s\) is the source impedance,  
\(Z_{\text{feedback}}(s)\) is the total feedback impedance, and  
\(G(s)\) is the frequency-dependent op-amp gain.

The analytical structure of the transfer function and feedback impedance follows established modeling and stability analysis approaches in~\cite{Chu2016, Goswami2018}, adapted to the voltage-sensing, current-injecting AEF topology shown in Figure~\ref{circuit}.

\paragraph*{Loop-gain}
The loop-gain that dictates phase- and gain-margin is

\begin{equation}
    T_{\text{feedback}}(s)=G_{OL}(s)\,\frac{G(s)\,Z_{\text{feedback}}(s)}{Z_{s}+Z_{\text{feedback}}(s)},
    \label{eq:loopgain}
\end{equation}

with \(G_{OL}(s)\) the intrinsic open-loop gain of the op-amp.  
A large \(\lvert T_{\text{feedback}}(j\omega)\rvert\) enhances noise attenuation but can reduce phase margin if not properly compensated.

\paragraph*{Feedback impedance}
From Figure \ref{circuit},

\[
Z_{\text{feedback}}(s)=
\Bigl(
  \tfrac{1}{R_{\text{feedback}}}
  +\tfrac{1}{R_{\text{comp}}+1/(sC_{\text{comp}})}
  +\tfrac{1}{R_{\text{comp1}}+1/(sC_{\text{comp1}})}
\Bigr)^{-1}.
\]

The resulting low- and high-frequency zeros flatten the phase response and suppress resonance.

\paragraph*{Bandwidth guideline}
Applying the decade-rule bandwidth guideline,

\begin{equation}
    f_{\text{cut-off}}
    =\frac{f_{\text{switching}}}{10^{\lvert\text{attenuation level}\rvert/40}},
    \label{eq:cutoff}
\end{equation}

ensuring the control bandwidth remains at least one decade below the switching frequency.

\paragraph*{RL-assisted tuning procedure}
The RL agent adjusts component values as follows:

\begin{itemize}
    \item \textit{Step 0 – Initialisation.}  
          Seed \(C_{\text{inject}}\), \(Z_{\text{damp}}\), and compensation values from the measured EMI spectrum.  
          Compute \(f_{\text{cut-off}}\) via Eq. \ref{eq:cutoff}.

    \item \textit{Step 1 – Injection-capacitor alignment.}  
          Fine-tune \(C_{\text{inject}}\) so its series resonance with \(L\) equals \(f_{\text{cut-off}}\):
          \begin{equation}
              L\,C_{\text{inject}}
              =\frac{1}{\kappa\,(2\pi f_{\text{cut-off}})^{2}},
              \label{eq:LCinject}
          \end{equation}
          where \(\kappa=C_{\text{sense}}/C_{\text{comp}}\) is the fixed capacitance ratio.

    \item \textit{Step 2 – Feedback-gain optimization.}  
          The agent iteratively updates \(G(s)\) for maximum attenuation while monitoring the phase margin.

    \item \textit{Step 3 – Damping-resistor tuning.}  
          With \(C_{\text{inject}}\) and \(\kappa\) fixed,
          \begin{equation}
              Z_\text{damp}
              =\kappa\,\sqrt{\frac{L}{\kappa\,C_{\text{inject}}}},
              \label{eq:Zdamp}
          \end{equation}
          to prevent low-frequency oscillations.

    \item \textit{Step 4 – Online fine-tuning.}  
          During operation, the policy continuously perturbs \(\{G(s), C_{\text{inject}}, Z_{\text{damp}}\}\) to maintain the optimal trade-off between attenuation and stability under varying EMI profiles.
\end{itemize}

\subsection{Problem Formulation and Representation of Decision Progress}
RL is a powerful branch of machine learning that enables adaptive control in dynamic environments. In this study, the AEF circuit with a tunable capacitor is modelled as an MDP, where the RL agent continuously interacts with the system to optimise capacitance and minimise EMI. The MDP is defined by a five-tuple ($S$, $A$, $P$, $\gamma$, $r$), where:
\begin{itemize}
    \item $S$ represents the set of all possible states within the AEF environment, consisting of EMI characteristics observed by the RL agent. At the timestep $t$, 
    \begin{equation}
    S_t = (\textit{EMI}_{\text{out}}, C_t,\textit{f}_{\text{response}})
    \end{equation}
    where $EMI_{\text{out}}$ is the measured EMI spectrum, $C_t$ is the capacitance value, and $f_{\text{response}}$ is the frequency domain response.
    \item $A$ represents the action space, which consists of discrete capacitor tuning adjustments. Since only the selected capacitor is tunable in this study, the RL agent can select an action that modifies the capacitance value, i.e.:
    \begin{equation}
    C_{t+1} = C_t + \Delta C
    \end{equation}
    where $\Delta C$ is capacitance adjustment chosen from a predefined set.
    \item $P(s_t, a_t \rightarrow s_{t+1})$ denotes the transition probability from state $s_t$ to state $s_{t+1}$ given action $a_t$.
    \item $\gamma \in (0, 1)$ is the discount factor that quantifies the value of future rewards compared to immediate rewards.
    \item $r(s,a)$ specifies the immediate reward received for taking action $a$ in state $s$, contributing to the total expected reward $\textbf{\textit{Reward}}$.
\end{itemize}

The cumulative long-term reward can be expressed as Eq. \ref{Equation 4}
\begin{equation}
    \textbf{\textit{Reward}} = \sum_{i=0}^{\infty} \gamma^i \, r \left( s_{t+i}, a_{t+i} \right)
    \label{Equation 4}
\end{equation}

To estimate the values of state-action pairs, the Bellman equation acts as a crucial link between theoretical concepts and practical implementation. It provides a recursive method for updating the Q-value, which represents the expected cumulative long-term reward for a given set of actions and states. The Q-function under the policy $\pi$ control is as shown in Eq. \ref{eq:Qdef}:
\begin{equation}
Q^{\pi}(s_t, a_t) = \mathbb{E}_{\pi}\!\left[\, r_t \,\middle|\, s_t, a_t \right],
\label{eq:Qdef}
\end{equation}

Where $Q^{\pi} (s,a)$ represents the cumulative rewards obtained by taking action $a$ in state $s$, and $\mathbb{E}_{\pi}$ is the expected return $\textbf{\textit{Reward}}$ under policy $\pi$. The expected value of the Q-function, averaged over all possible actions under a given policy, constitutes the state value function $\textbf{\textit{Value}}^{\pi} (s)$. This value can be calculated by taking the expectation of the Q-function across the entire action space $A$,  as mathematically expressed in Eq. \ref{Equation 6}:

\begin{equation}
    \textbf{\textit{Value}}^{\pi}(s_t) = \sum_{a_t \in A} \pi(a_t \mid s_t)\, Q^{\pi}(s_t, a_t).
    \label{Equation 6}
\end{equation}
Eq. \ref{Equation 6} is formulated for a stochastic polic
$\pi(a_t\mid s_t)$. 
In the process of tuning AEF circuit parameters, the state space reflects observations from the environment, capturing a comprehensive set of measured data. This includes sensed interference amplitudes, filter insertion losses, current, voltage, and overall filtering performance under controlled conditions. These parameters enable a precise understanding of and adjustment to the circuit’s operational dynamics. Thus, the relationship between state and action at time $t$ should be represented by the sensed information from the AEF circuit $s_t$. The adjustment $a_t$ is made based on $s_t$, and after applying $a_t$, the updated circuit cancellation performance $s_{t+1}$ is realized. 

In tuning the AEF circuit parameters, the state space reflects observations from the environment and captures a comprehensive set of measured data (e.g., interference amplitudes, filter insertion losses, current, voltage, and overall filtering performance). These variables enable precise understanding and adjustment of the circuit’s operational dynamics. At time step $t$, the state is denoted by $s_t$. Based on $s_t$, the agent selects an action $a_t$; after applying $a_t$, the environment transitions to the next state $s_{t+1}$ with an updated EMI response. The reward function in Eq.~\eqref{eq:reward} encourages the agent to make optimal choices.
\begin{equation}
r_t =
\begin{cases}
\log_{10}(EMI_{s_t}) - \log_{10}(EMI_{s_{t+1}}), 
& \text{if } EMI_{s_{t+1}} \le \tau_{\mathrm{EMI}},\\\\
-15, & \text{otherwise.}
\end{cases}
\label{eq:reward}
\end{equation}
Here, $EMI_{s_t}$ and $EMI_{s_{t+1}}$ denote the measured EMI output intensity before and after executing the tuning action at time step $t$, respectively. Specifically, $EMI_{s_t}$ corresponds to the EMI level observed at state $s_t$, whereas $EMI_{s_{t+1}}$ denotes the filtered EMI level obtained at the next state $s_{t+1}$ after applying $a_t$. The term $\tau_{\mathrm{EMI}}$ is the target EMI threshold that defines the acceptable emission level, i.e., 15 dB$\mu$A in this study. If $EMI_{s_{t+1}} \le \tau_{\mathrm{EMI}}$, the reward equals the achieved attenuation (on a logarithmic scale); otherwise, a constant penalty of $-15$ discourages unstable or ineffective tuning behaviours.

To ensure that the final strategy exhibits optimal generalization capabilities in offline applications and can accurately adapt to new environments for precise filtering outcomes, we focused on minimizing the agents’ reliance on stable circuit operational conditions. Additionally, we aimed to improve their robustness against variations in circuit parameters and external disturbances. To achieve this, we propose using a VAE model to process observational data from circuits and their corresponding responses. The goal is to learn compact representations of this data and the associated actions in a latent space.

\begin{figure}[h]
\centering
\includegraphics[width=0.485\textwidth]{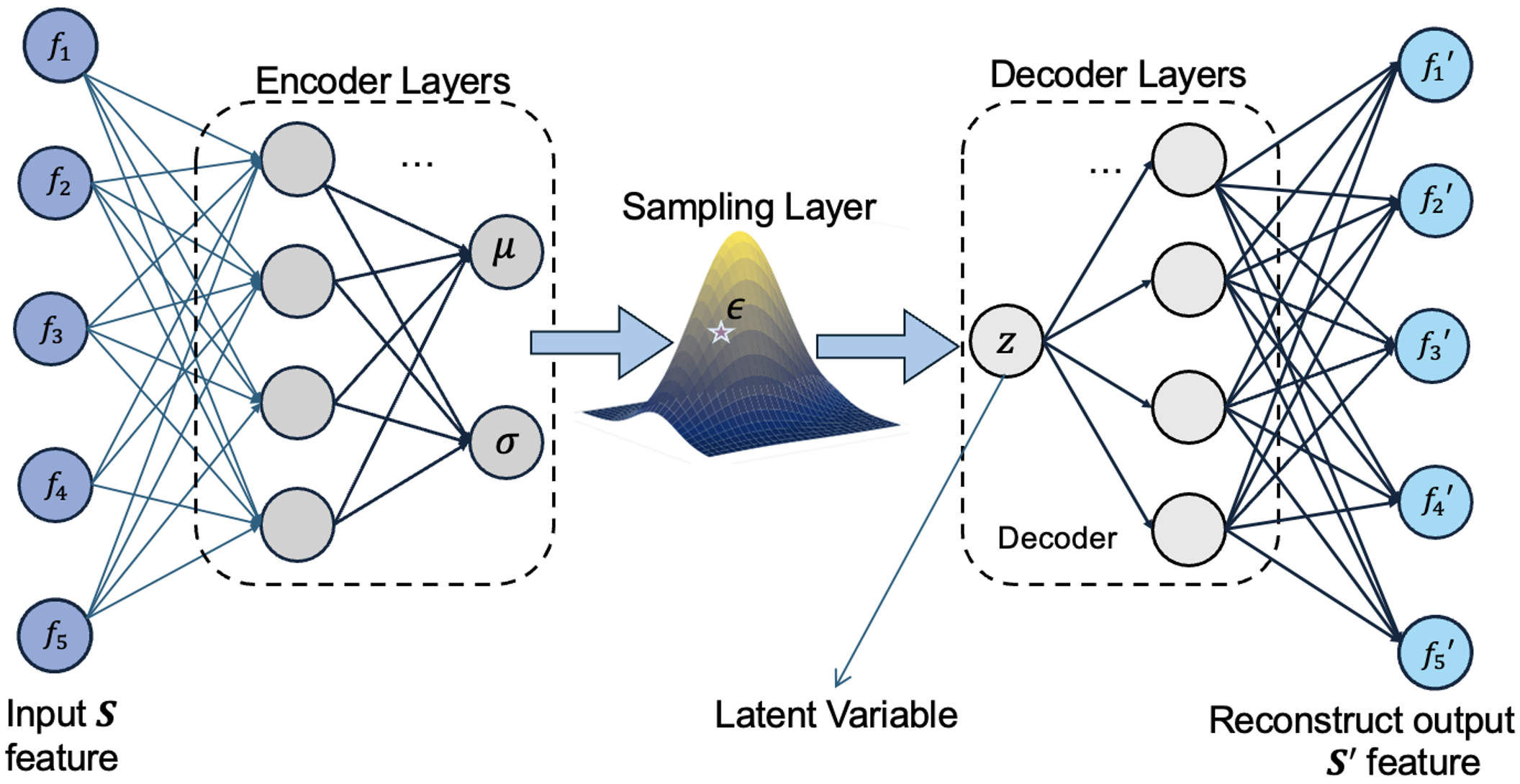}
\captionsetup{justification=centering} 
\caption{The structure of VAE: the left-side encoder maps input to the parameters from a Gaussian distribution while the right-side decoder reconstructs the latent variable from the sampling layer as output data}
\label{VAE}
\end{figure}

The model consists of a deep encoder-decoder architecture, as shown in Figure \ref{VAE}. During the encoding phase, the structure incorporates three layers of a Multilayer Perceptron (MLP), combined with the nonlinear activation function Rectified Linear Unit (ReLU) and fully connected layers, to extract critical features from the observational circuit data. Unlike traditional autoencoders, the VAE encodes the data into a probabilistic latent space, where the encoder outputs parameters, as depicted in Figure 6. Specifically, the means  $\mu$ and standard deviations  $\sigma$ follow an assumed Gaussian distribution to represent latent variables $z \mid s \sim \mathcal{N}(\mu, \sigma^2)$, as expressed by Eq. \ref{Equation 8}. These probabilistic representation captures features such as interference types, intensity levels, and response patterns within the circuits, providing a robust foundation for modelling variability and noise.

\begin{equation}
    z = \mu + \sigma \cdot \epsilon
    \label{Equation 8}
\end{equation}
where $\epsilon$ represents a noise term sampled from standard deviations $\mathcal{N}(0, 1)$, facilitating the successful application of backpropagation.

\begin{figure}[h]
\centering
\includegraphics[width=0.485\textwidth]{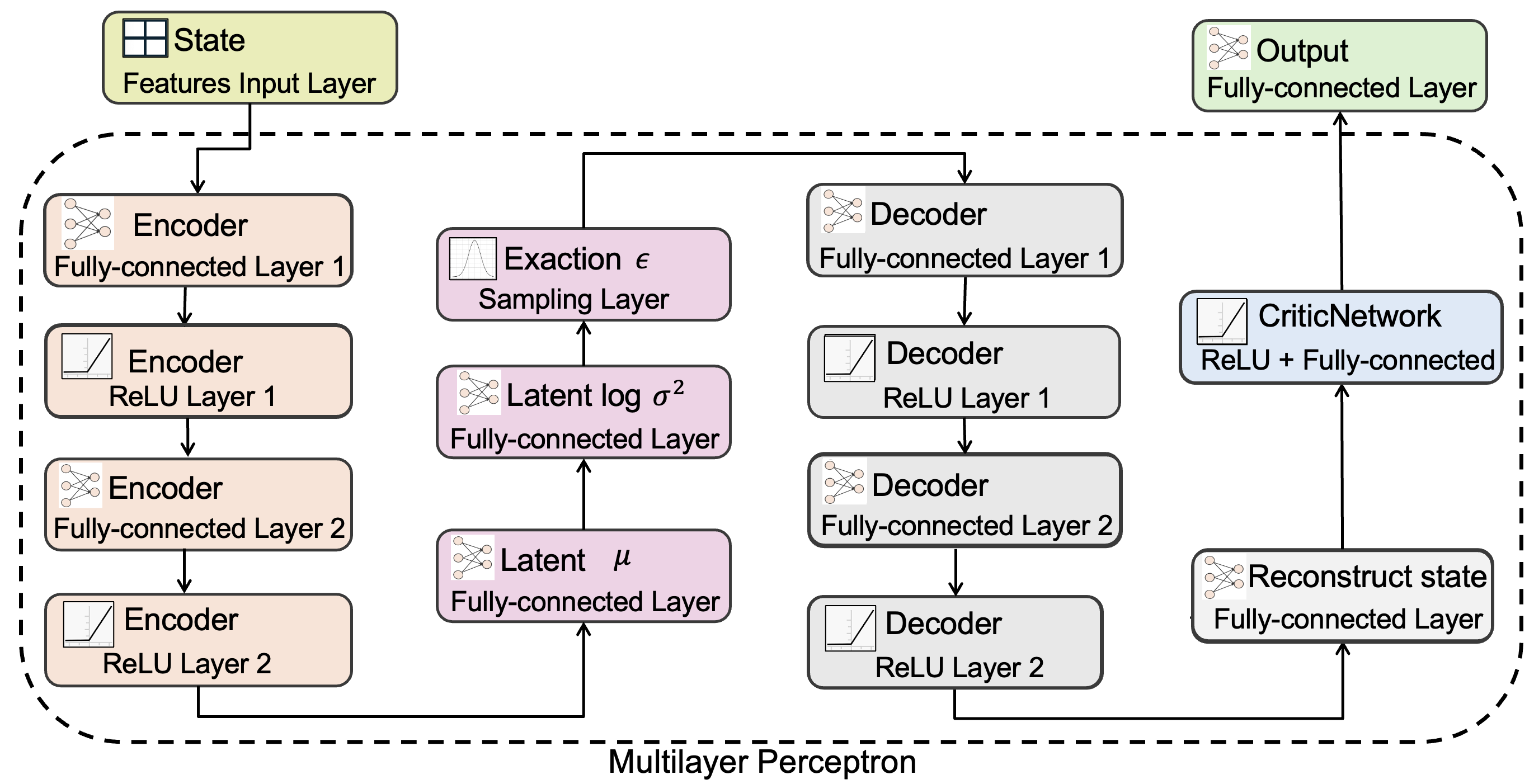}
\captionsetup{justification=centering} 
\caption{The framework of the MLP constitution}
\label{MLP}
\end{figure}

Sampling from these distributions is accomplished using the reparameterization technique, which enables backpropagation through stochastic operations to facilitate learning. The sampled latent representation is then processed by a decoder composed of four MLP layers to reconstruct outputs that closely resemble the original data. This reconstruction validates and reinforces the intrinsic characteristics of the data, as depicted in Figure \ref{MLP}.

Following the feature extraction and data encoding process, the output layer of the MLP is directly linked to the estimation of action values. The agent uses the observed states to predict the expected values for each action and selects actions based on these predictions. Simultaneously, it updates the MLP’s weights using rewards and new state information obtained from the environment. This end-to-end structure not only functions as a feature extractor but also serves as an approximator for the value function, enabling the agent to make rapid and accurate decisions in complex environments. By employing this approach, we can enhance the adaptability and performance of the strategy, ensuring that the agent maintains stable and efficient operations, even in the face of environmental changes and disturbances. This configuration demonstrates excellent generalization capabilities, allowing the agent to adapt accurately to new environments and provide precise filtering effects.

\subsection{Fundamentals of Proposed Enhanced Algorithm}
Our objective is to optimize and stabilize the real-time EMI cancellation response of AEF circuits while minimizing dependency on environmental changes. To achieve this, we adopt a model-free, Q-value-based algorithm as a reliable method. To further enhance exploration efficiency, in addition to using the VAE with a Gaussian distribution for optimized state representation, we also plan to incorporate a noise network into the action selection process during policy execution. This prevents the strategy from prematurely converging on suboptimal solutions, as described in Eq. \ref{Equation 9}. The increased noise disturbance factor $\eta$ facilitates broader exploration, allowing the agent to explore the potential policy space more thoroughly.
\begin{equation}
    a_{t+1} = \arg \max_{a} \left[ Q(s_{t+1}, a_{t}) + \eta \right]
    \label{Equation 9}
\end{equation}
Furthermore, we utilize the temporal difference (TD) error to update the Q-values, incorporating momentum and adaptive learning rates to improve the stability and efficiency of the learning process. This approach promotes convergence towards the target Q-values, as outlined in Eqs. \ref{Equation 10} and \ref{Equation 11}.
\begin{equation}
    Q(s_t, a_t) \leftarrow Q(s_t, a_t) + \alpha \left[ Q_{\text{target}} - Q(s_t, a_t) \right]
    \label{Equation 10}
\end{equation}
\begin{equation}
    Q_{\text{target}} = r_t + \gamma \cdot Q(s_{t+1}, a_{t+1}; \text{critic network})
    \label{Equation 11}
\end{equation}
To effectively monitor the agent’s progress in optimizing its strategy, it is crucial to evaluate its performance throughout the training process. To this end, we establish a loss function, as defined in Eq. \ref{Equation 12}, which evaluates the updates made to the policy by minimizing the loss at each iteration. This approach not only tracks the learning progress but also ensures that the agent continuously refines its behaviour during the training phase.
\begin{equation}
    \textbf{\textit{Loss}} = \mathbb{E} \left[ \left( r_t + \gamma \max_{a_{t+1}} Q(s_{t+1}, a_{t+1} \mid \theta) - Q(s_t, a_t \mid \theta) \right)^2 \right]
    \label{Equation 12}
\end{equation}
Here, $\gamma$ is the discount factor, and $max_{a_{t+1}} Q(s_{t+1}, a_{t+1} \mid \theta)$ is the estimate $Q$ value for the next state $s_{t+1}$ and action $a_{t+1}$. The parameter $\theta$ represents the weight of the MLP. 

\begin{figure}[h]
\centering
\includegraphics[width=0.485\textwidth]{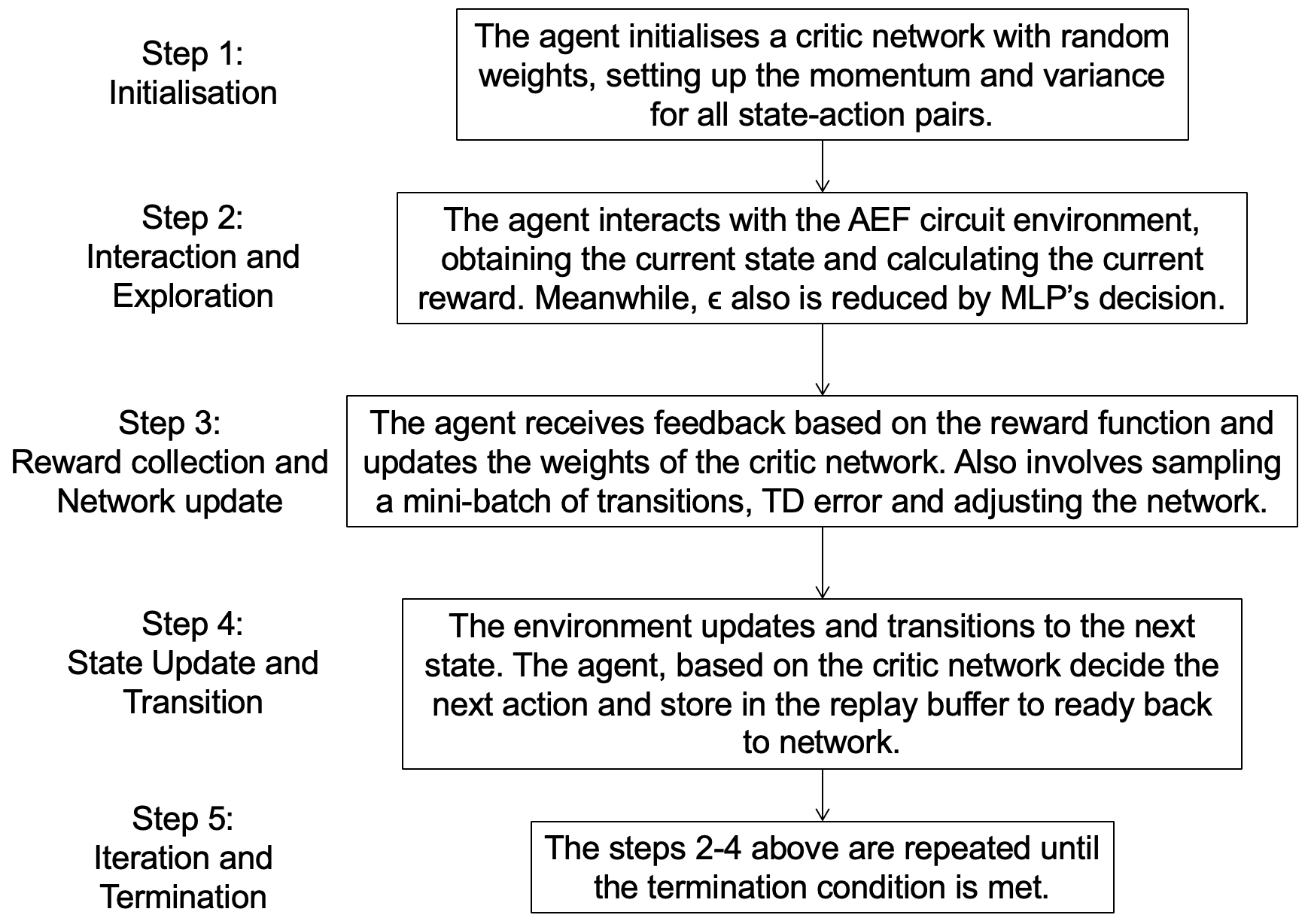}
\captionsetup{justification=centering} 
\caption{The flowchart of the proposed EQRL algorithm}
\label{EQRL}
\end{figure}

The RL-assisted tuning process consists of two distinct phases: a training phase and an execution (inference) phase. During the training phase, the EQRL agent continuously interacts with the Simulink-based self-tuning AEF environment. At each iteration, the agent observes the current EMI state, selects an action (e.g., adjustment of injection capacitance or damping impedance), and receives a reward determined by the EMI attenuation performance. Algorithm 1 governs this training loop, in which both the Q-values and the neural-network parameters are iteratively updated until convergence to an optimal control policy. Once training is completed, the learned policy parameters are fixed. In the subsequent execution phase, the agent no longer performs learning or parameter updates but directly applies the trained policy to output control actions for new EMI states, thereby enabling real-time adaptive tuning of the AEF system. The EQRL algorithm is outlined in Algorithm 1. The working steps involved in the proposed EQRL are summarised in Figure \ref{EQRL}.

\begin{algorithm}[h!]
\caption{Enhanced Q-value based RL (EQRL) algorithm for self-tuning AEF circuit}

Initialize critic network with random weights\;
Initialize $Q(s,a)=0$ and $\pi(s,a)=0$ for all $s,a$\;
Initialize replay buffer $D$ to store tuples $(s_t,a_t,r_t,s_{t+1})$\;
Observe initial state $s_0$\;
Select initial action $a_0$ randomly from action space $A$\;
Set $maxEpisode$\;

\For{episode $=1$ \textbf{to} $maxEpisode$}{
    $t=0$\;
    \While{$s_t$ is not terminal}{
        Execute $a_t$ and observe next state $s_{t+1}$\;
        
        \If{$EMI_{t+1} \le \tau_{\mathrm{EMI}}$}{
            $r_t = \log_{10}(EMI_s)-\log_{10}(EMI_{t+1})$\;
        }\Else{
            $r_t = -15$\;
        }

        Store $(s_t,a_t,r_t,s_{t+1})$ in $D$\;

        \If{rand() $< \varepsilon$}{
            Select $a_{t+1}$ randomly from $A$\;
        }\Else{
            $a_{t+1}=\arg\max_a Q(s_{t+1},a_{t})$, using critic network\;
        }

        \If{length($D$) $\ge$ BatchSize}{
            Sample a random minibatch of transitions from $D$\;
            \For{each $(s_t,a_t,r_t,s_{t+1})$ in minibatch}{
                \uIf{$s_{t+1}$ is terminal}{
                    $Q_{\text{target}} = r_{t}$\;
                }\Else{
                    $Q_{\text{target}} = r_{t} + \gamma Q(s_{t+1},a_{t+1}; \text{criticNetwork})$\;
                }

                Calculate loss: 
                $L = (Q_{\text{target}} - Q(s_{t+1},a_{t+1}; \text{criticNetwork}))$\;

                $m(s,a)=\beta_1 m(s,a)+(1-\beta_1)\delta$\;
                $v(s,a)=\beta_2 v(s,a)+(1-\beta_2)\delta^2$\;
                $\hat{m}=\frac{m_t(s,a)}{1-\beta_1^{t+1}}$\;
                $\hat{v}=\frac{v_t(s,a)}{1-\beta_2^{t+1}}$\;

                $Q(s,a)=Q(s,a)+\alpha\frac{\hat{m}}{\sqrt{\hat{v}}+\varepsilon}+\eta$\;
            }
        }

        $s_t\leftarrow s_{t+1}$;\quad $a_t\leftarrow a_{t+1}$\;
        $\varepsilon=\max(\varepsilon\cdot0.95,0.1)$\;
        $t\leftarrow t+1$\;
    }
}
\end{algorithm}

\section{Experimental Setup}
This section outlines the implementation of the simulation environment and the algorithmic framework used in this study. Both the proposed algorithm and the simulator were developed and executed using the MATLAB/Simulink 2024 RL Toolbox. The simulations were conducted on a system equipped with an AMD Ryzen 7 5800H processor featuring integrated Radeon Graphics, operating at a clock speed of 3.20 GHz. The system operates on a 64-bit platform supported by an x64-based processor architecture.

\subsection{Data Processing}
In this study, the EMI signal dataset was collected from the output cables of an EDU. Table \ref{Table 1} summarizes the dataset’s composition and frequency range. To ensure precise capture and analysis of the EMI noise data, a detailed numerical analysis was performed. The noise information was extracted from the cables, and the data processing procedure involved generating sinusoidal waves corresponding to each detected EMI frequency and its harmonics. These generated signals were then introduced into the simulation circuit as a time series, emulating real-world EMI conditions. Following this, adaptive noise cancellation filtering was applied using the AEF. The resulting filtered data were subsequently transformed into the frequency domain via Fast Fourier Transform (FFT) for further performance analysis.

\begin{table}[h!]
\centering
\caption{Description of datasets and their characteristics}
\begin{tabular}{p{1.25cm} p{0.75cm} p{3.5cm} p{1.5cm}}
\toprule
\textbf{Datasets} & \textbf{EMI range} & \textbf{Input Variables} & \textbf{Interferences amount} \\
\midrule
Cable 44 & 150k-30MHz & Frequencies, bandwidth, peak values, average values, minimum values, time intervals. & 6650 \\
Cable 47 & 100k-30MHz & Frequencies, bandwidth, peak values, average values, minimum values, time intervals. & 13290 \\
Cable Bundle & 150k-30MHz & Frequencies, bandwidth, peak values, average values. & 13290 \\
\bottomrule
\end{tabular}
\label{Table 1}
\end{table}

In the signal analysis process, a Hanning window is applied to mitigate spectral leakage, ensuring smoother signal boundaries. This windowing function allows the signal processed by the FFT to approximate a periodic signal, providing a more accurate representation of the true frequency components in the spectrum. Eq. \ref{Equation 13} presents the formula for calculating the output of the filtered signal, where $Magnitude (f)$ represents the signal amplitude, $N$ denotes the number of samples in the signal, $x(n)$ is the discrete signal value, and $\omega(n)$ refers to the window function. This method allows for precise quantification of the attenuation performance of the AEF against EMI. It is particularly useful in the tunable AEF circuit, which is controlled by an RL algorithm. To validate the feasibility of the proposed RL-assisted tunable solution for EMI suppression in the circuit, the topology shown in Figure \ref{circuit} is employed as the operational environment for both the input signal and the resulting output data.

\begin{equation}
    \textit{Magnitude} (f) = \frac{2}{N} \cdot \left| \text{FFT} \left[ x(n) \cdot \omega(n) \right] \right|
    \label{Equation 13}
\end{equation}

To emulate real-world EMI conditions within the simulation environment, EMI signals were injected into the AEF circuit model through the power line input using synthetic waveforms derived from measured EMI spectra. Specifically, sinusoidal signals representing dominant frequency components and their harmonics, extracted from the Cable Bundle dataset, were superimposed on the (direct current) DC input of the EDU simulation block. These signals were constructed using a time-series waveform generator in MATLAB and fed into Simulink via signal source blocks. The RL agent, implemented using the MATLAB RL Toolbox, interfaces with Simulink through an environment wrapper that exposes tunable parameters (e.g., $C_{\text{inject}}$) and circuit state variables (e.g., EMI amplitude, frequency response). At each timestep, the agent observes the EMI-related states from the Simulink model, selects an action to adjust the tunable parameter value, and receives a reward based on the resulting attenuation, forming a closed-loop co-simulation between the RL logic and circuit dynamics.

\subsection{Tunable Circuit Simulation and Training}
\label{sec:Tunable Circuit Simulation and Training}
\begin{figure}[h]
\centering
\includegraphics[width=0.485\textwidth]{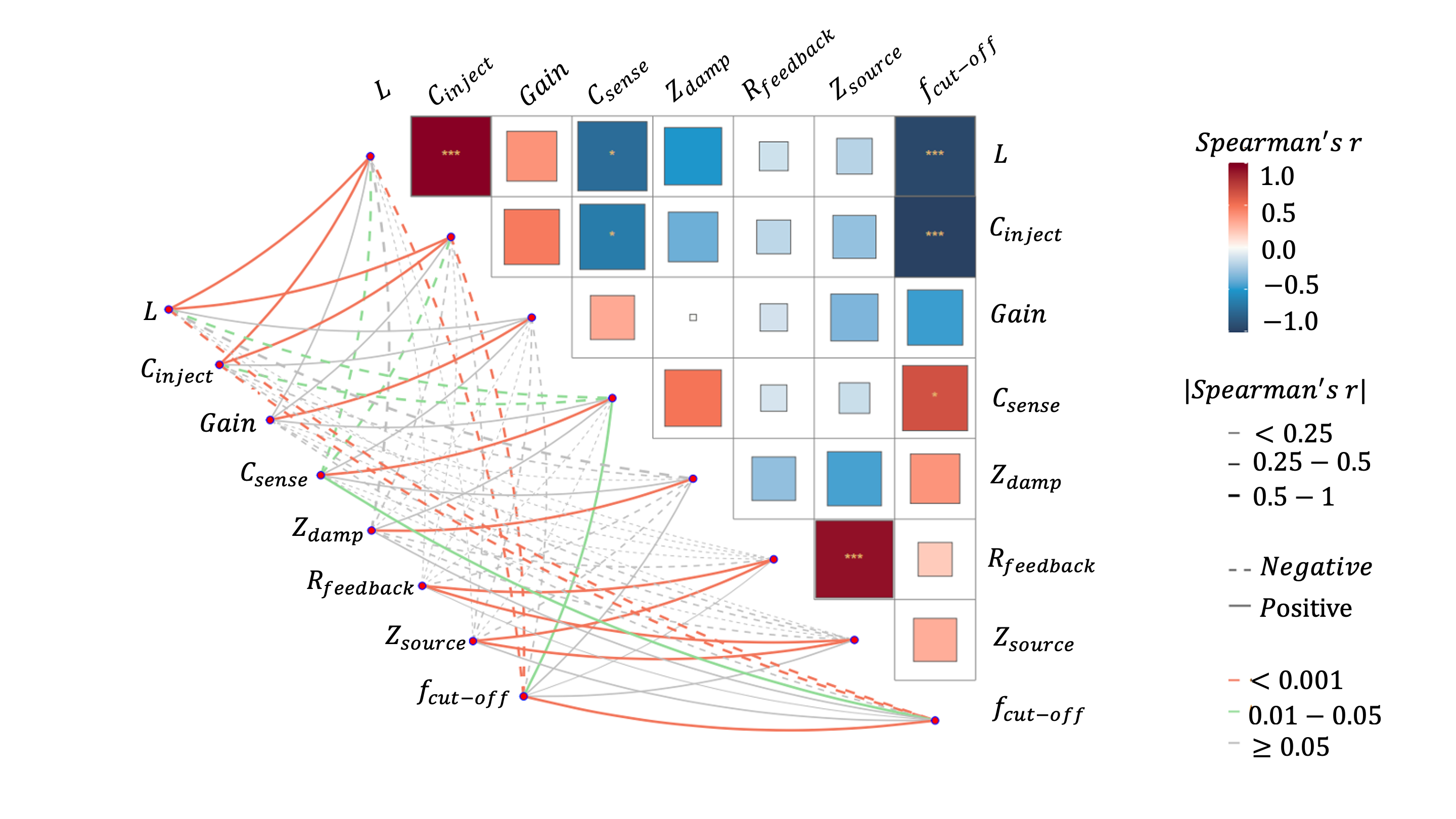}
\captionsetup{justification=centering} 
\caption{Spearman correlation heatmap and significance analysis}
\label{heatmap}
\end{figure}

The selection of the most effective tunable component is crucial for achieving the optimal control objective in the tunable model. Figure \ref{heatmap} presents a Spearman correlation heatmap, accompanied by a significance analysis, which illustrates the relationships between various key components and their influence on the cut-off frequency. Through this analysis, it was determined that the injection capacitor and inductor have the most significant impact on filter performance. Specifically, adjustments to these components result in notable changes to the cut-off frequency. However, in practical circuits, the physical size of inductors, combined with parasitic capacitance and resistance at high frequencies, can lead to instability in inductance values, which negatively affects circuit performance. In high-frequency applications, these parasitic effects can cause shifts in resonant frequency, complicating tuning efforts. As a result, the injection capacitor is identified as the most suitable component for self-tuning. By employing an RL policy to control the injection capacitor, the AEF can optimize its performance in dynamic EMI environments, effectively adapting to varying noise sources and enabling precise, real-time EMI suppression.

A key concern regarding the feasibility of inductor-based tuning in EMI filter applications is the challenge posed by dynamic inductance adjustments due to the physical nature of inductors. Although solutions such as switchable inductance networks and magnetic core biasing offer limited adjustability, they also introduce added complexity, size, and switching noise. In contrast, active filtering techniques or field-programmable analogue arrays provide more practical alternatives by digitally emulating inductance behaviour. Therefore, while real-time adjustment of physical inductors remains complex and less practical for high-frequency EMI filters, the use of injection capacitors for tuning with a set tunable range, controlled by the RL policy, offers a more stable and precise method for dynamic EMI suppression. By utilising the RL policy to manage the injection capacitor, the AEF achieves optimal performance in dynamic EMI environments, adapting to varying noise sources and enabling precise and efficient EMI suppression. This approach ensures that the filter can maintain high performance without the limitations associated with inductors.

The AEF circuit parameters, listed in Table \ref{Table 2}, define the key operational settings of the active EMI filter. These include components such as inductors, capacitors, and resistors, all of which directly influence critical performance factors, including the cut-off frequency, filter bandwidth, and the overall effectiveness of EMI suppression. Properly configuring these parameters ensures that the RL-based tuning strategy operates within its optimal range, providing the necessary flexibility to adapt to dynamic electromagnetic interference environments and effectively suppress them.

\begin{figure}[h]
\centering
\includegraphics[width=0.485\textwidth]{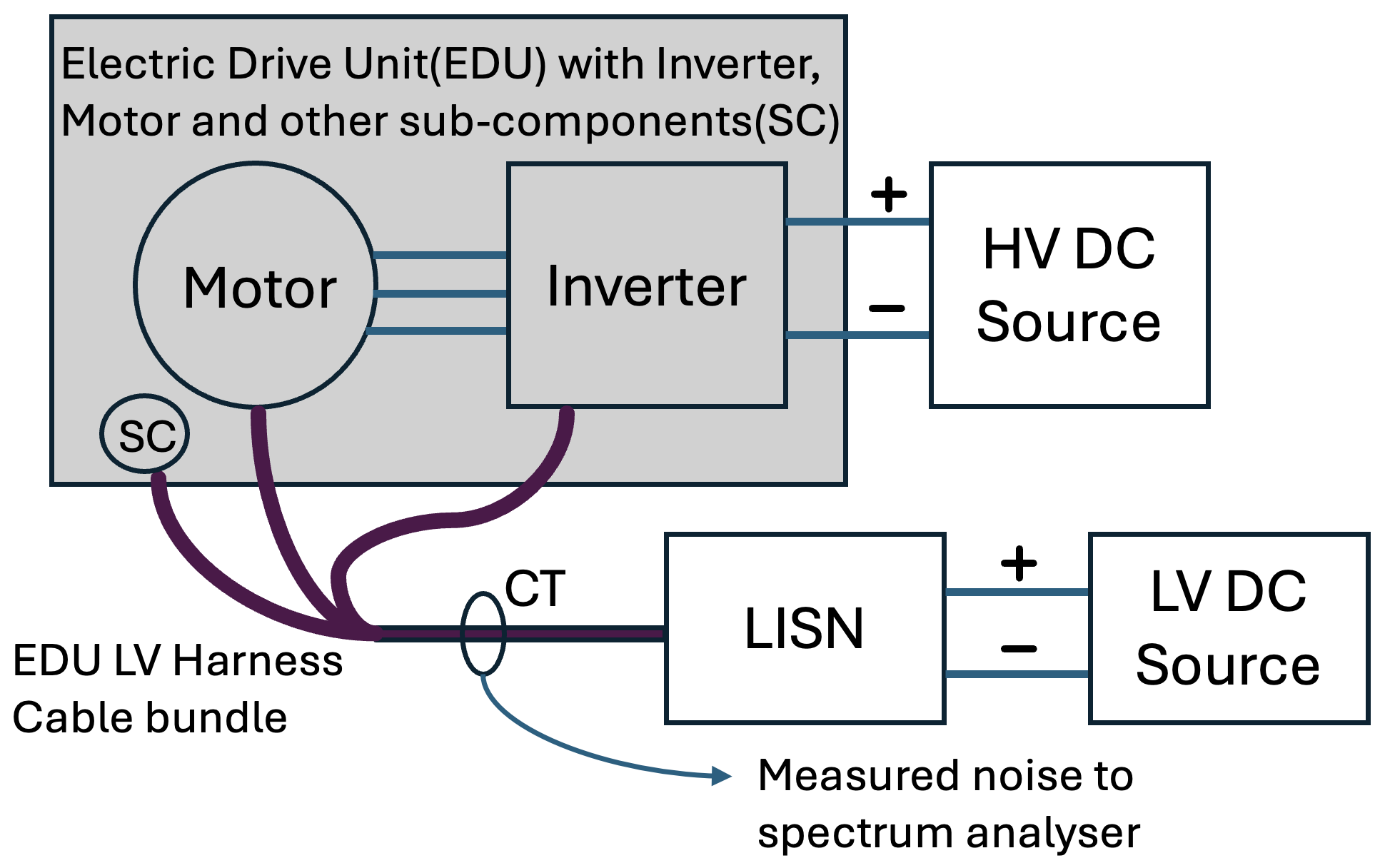}
\captionsetup{justification=centering} 
\caption{Noise measurement from EDU LV harness}
\label{measurement}
\end{figure}

In this work, we have used practical measurement data from an EDU to train the RL algorithm and adjust the active EMI filter parameters. The measurement block diagram from the Low Voltage (LV) harness cable bundle is shown in Figure \ref{measurement}.  

In this figure, EMI noise is detected from the LV harness using a high-bandwidth current transformer (CT). The LV harness comprises cables originating from the Inverter, Motor, and other active sub-components of the EDU. The Line Impedance Stabilisation Network (LISN) lies between the LV DC power supply and the component side of the harness. In the present work, the proposed active filter is being developed in a simulation platform, and we are currently in the process of creating the first prototype of the proposed active filter as the next phase of this project. The active EMI filter schematic is depicted in Figure \ref{circuit} The related component values were chosen based on established principles in AEF design, as outlined in the methodologies proposed by Chu et al. \cite{Chu2022, Chu2016}. The key parameters, such as cut-off frequency, injection capacitance, feedback resistance, and damping resistance, were identified using standard filter design equations within these methodologies. The values presented in Table \ref{Table 2} were derived from \cite{Chu2022} and are widely regarded as optimal for designing AEFs that effectively suppress EMI and maintain stability. Although the injection capacitance $C_{\text{inject}}$ is physically continuous, the EQRL implementation employs a discrete action space by dividing the tuning range (10 pF–500 nF) into a small symmetric set of incremental steps (e.g., $\pm \Delta C$, $\pm 2\Delta C$) around the nominal value to ensure stable Q-value updates and efficient training.

\begin{table}[h!]
\centering
\caption{Components parameters for the AEF circuit}
\begin{tabular}{ll}
\toprule
\textbf{Item} & \textbf{Value} \\
\midrule
$C_{sense}$ & 100 nF \\
$C_{inject}$ & 470 nF (basic) \\
             & (\textit{RL Control range} 10 pF - 500 nF) \\
$C_{comp}$ & 1 nF \\
$C_{comp1}$ & 100 nF \\
$L$ & 1 $\mu$H \\
$R_{comp}$ & 1 k$\Omega$ \\
$R_{comp1}$ & 0.5 $\Omega$ \\
$R_{feedback}$ & 50 k$\Omega$ \\
$Z_{damp}$ & 1.8 $\Omega$ \\
$Z_{source}$ & 5 $\Omega$ \\
\bottomrule
\end{tabular}
\label{Table 2}
\end{table}

In addition to the circuit parameters, the RL training hyperparameters, outlined in Table \ref{Table 3}, play a pivotal role in shaping the learning process of the agent. These hyperparameters include the learning rate, discount factor, exploration-exploitation ratio, and batch size, each of which significantly influences how efficiently the agent learns to tune the AEF. For instance, the learning rate determines how quickly the agent updates its policy in response to feedback from the environment, while the exploration-exploitation ratio balances the exploration of new actions and the exploitation of known successful strategies. The selection and fine-tuning of these hyperparameters were guided by iterative experimentation and previous research, ensuring that the agent can learn an effective policy within a reasonable time frame, while avoiding overfitting to the training data.

\begin{table}[h!]
\centering
\caption{Hyperparameters setup for algorithms}
\begin{tabular}{lp{4cm}}
\toprule
\textbf{Variable} & \textbf{Value} \\
\midrule
Learning rate & 0.001 \\
Discount factor & 0.99 \\
Max episode & 50 \\
Optimizer & Adam \\
Noise disturbance factor & 0.025 \\
Max step size & 6000 \\
\midrule
\textit{Q-learning} & \\
Epsilon greedy Exploration & 1/0.001/0.5 \\
Critic network & (400, 300) \\
\midrule
\textit{SARSA} & \\
Epsilon greedy Exploration & 1/0.001/0.5 \\
\midrule
\textit{DQN} & \\
Replay buffer size & 100000 \\
Epsilon greedy Exploration & 0.95/0.001/0.5 \\
Target update rate & 4 \\
\midrule
\textit{The proposed algorithm} & \\
Epsilon greedy Exploration & 0.99/0.001/0.5 \\
Greedy decay rate & 0.005 \\
Replay buffer size & 100000 \\
Critic network & (128, 64, 2, 2, 64, 128, 80) \\
\bottomrule
\end{tabular}
\label{Table 3}
\end{table}

Table \ref{Table 4} outlines the dataset configuration for RL training and testing. The policy is trained through real-time interaction with the Cable Bundle dataset, which serves as the primary training environment where the agent interacts directly with the circuit. Additional datasets are used for offline testing to verify the policy’s generalization capability and robustness. The training process begins by applying a deep RL-based approach to optimize the tuning of $C_{\text{inject}}$ from its initial improper value, allowing the agent to dynamically adjust the AEF’s cut-off frequency in response to EMI variations. Empirical testing indicated that a training duration of 50 episodes was sufficient to stabilize the learning process, enabling robust policy outcomes. Consequently, the training was limited to 50 episodes, a decision substantiated by the results, which demonstrated effective policy learning within this timeframe. The agent continuously interacts with the environment, updating its policy based on the observed EMI characteristics and the corresponding filter performance. Over time, this interaction enables the agent to learn an optimal tuning strategy that minimizes EMI across a wide range of conditions, ultimately improving filter performance in real-world applications.

\begin{table}[h!]
\centering
\caption{Datasets arrangement for training and testing}
\begin{tabular}{lccc}
\toprule
\textbf{Datasets} & \textbf{Train} & \textbf{Extraction test} & \textbf{Offline test} \\
\midrule
Cable 44 & $\times$ & $\times$ & 6650 \\
Cable 47 & $\times$ & $\times$ & 6769 \\
Cable Bundle & 13290 & 4430 & $\times$ \\
\bottomrule
\end{tabular}
\label{Table 4}
\end{table}

\subsection{Evaluation Metrics}
To accurately evaluate the performance of the RL algorithm in controlling AEF parameters, several key metrics are used to assess both the effectiveness and accuracy of the algorithm. The primary metric is the insertion loss, measured in decibels (dB), which quantifies the EMI suppression achieved by the AEF under RL control. The insertion loss can be calculated using Eq. \ref{Equation 14}, which expresses a logarithmic ratio of the EMI intensity before and after filtering. This metric is widely employed in EMI filtering applications to assess how effectively unwanted interference is attenuated.

\begin{equation}
    \textit{Insertion loss (dB)} = 20 \log_{10} \frac{I_o}{I_f}
    \label{Equation 14}
\end{equation}
where $I_o$ represents the original EMI intensity, and $I_f$ denotes the filtered EMI intensity.
In addition to the insertion loss, the Root Mean Squared Error (RMSE) is used to assess the accuracy of the AEF’s performance when controlled by the RL algorithm. The RMSE metric compares the insertion loss achieved under RL control with the required insertion loss across multiple test scenarios. This is essential for verifying the algorithm’s generalization capabilities and consistency of performance across varying EMI environments. The RMSE can be calculated using Eq. \ref{Equation 15}:

\begin{equation}
    RMSE = \sqrt{ \frac{1}{N_g N_p} \sum_{i=1}^{N_g} \sum_{j=1}^{N_p} \left[ G_{a,i}(f_j) - G_{r,i}(f_j) \right]^2 }
    \label{Equation 15}
\end{equation}
where $N_g$ represents the number of unknown scenarios test dataset groups, and $N_p$ is the number of EMI data points within each test dataset group. $G_{a,i}(f_j )$ is the actual needed insertion loss for the $i$-th test dataset group at the $j$-th EMI data point, and $G_{r,i}(f_j)$ is the insertion loss of of $i$-th test dataset group at the $j$-th EMI data point under RL controlled performance.

\section{Results and Discussion}
In Sections \ref{sec:Self-tuning AEF System Design} and \ref{sec:Tunable Circuit Simulation and Training}, the self-tuning AEF model, controlled by an RL policy, was designed and simulated, respectively. All experimental comparisons were conducted using EMI noise levels sourced from the dataset outlined in Table \ref{Table 1}, which covers a bandwidth ranging from 100 kHz to 30 MHz. This section presents a comparative analysis of training performance across various algorithms, with a primary focus on the proposed EQRL algorithm.

\subsection{Comparative Analysis of Training Performance}
\label{sec:Comparative Analysis of Training Performance}
In this section, we conducted a comparative evaluation of the training performance of various RL algorithms, with a particular focus on our proposed EQRL algorithm. The training process aimed to optimize the tunable components of the adaptive AEF circuit to achieve the desired EMI suppression. To assess the effectiveness of each algorithm, two key aspects were evaluated: convergence speed and cumulative reward, both of which are critical indicators of learning efficiency.

Firstly, convergence speed refers to the time required for each algorithm to consistently achieve optimal filtering results. The faster the convergence, the more time-efficient the algorithm is in dynamically adjusting circuit parameters. Secondly, the cumulative reward serves as a quantitative measure of the agent’s learning process, indicating how effectively the algorithm maximizes the reward function, which, in this case, is directly linked to the filter’s EMI suppression performance. As shown in Table \ref{Table 5}, the experimental results highlight significant differences in convergence speed and overall learning efficiency across the tested algorithms. Notably, the EQRL algorithm demonstrated a faster convergence time compared to traditional algorithms such as State–action–reward–state–action (SARSA) and standard Q-learning. This rapid convergence is particularly advantageous in real-time applications, where quickly training an optimal policy can lead to substantial savings in both time and human resources.

\begin{table}[h!]
\centering
\caption{Quantitative rewards}
\begin{tabular}{p{1.75cm} p{3cm} p{2.5cm}}
\toprule
\textbf{Algorithms} & \textbf{Cumulative average reward} & \textbf{Running time/episode} \\
\midrule
SARSA & -7638.80 $\pm$ 926.76 & 86.08 s \\
Q-Learning & -7823.40 $\pm$ 505.93 & 81.40 s \\
DQN & -9628.80 $\pm$ 1894.49 & 11.16 s \\
EQRL & -5601.86 $\pm$ 501.53 & 10.07 s \\
\bottomrule
\end{tabular}
\label{Table 5}
\end{table}

\begin{figure}[h]
\centering
\includegraphics[width=0.485\textwidth]{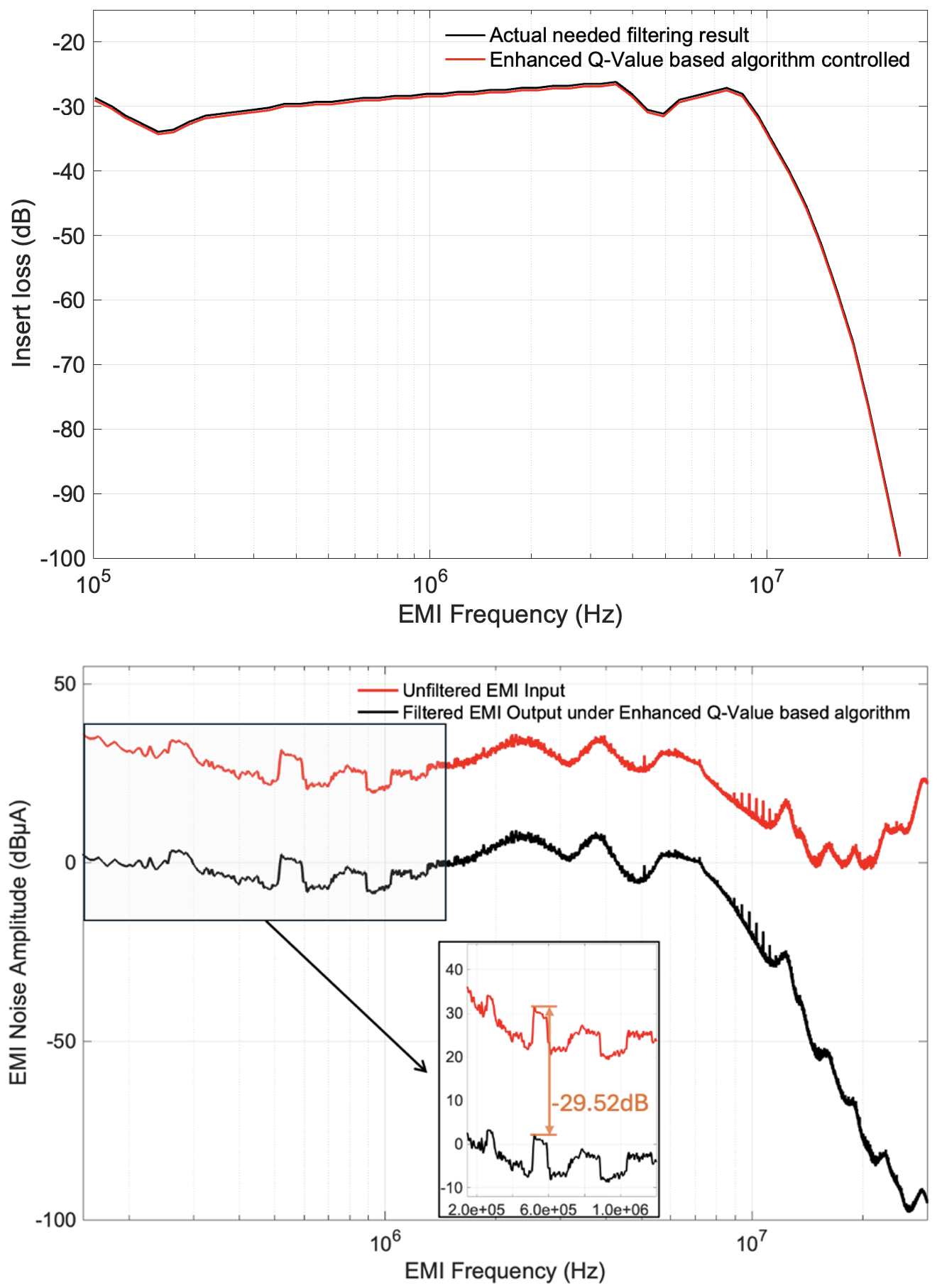}
\captionsetup{justification=centering} 
\caption{Extraction cable bundle dataset insertion loss and EMI signal filtering performance}
\label{extraction}
\end{figure}

Additionally, the cumulative reward values clearly show that the EQRL algorithm consistently outperformed the other algorithms, including Deep Q-Network (DQN), in optimizing the AEF circuit parameters. Maintaining higher average reward scores throughout the training process proved more effective in achieving the desired filtering performance, as evidenced by the close alignment between the actual required filtering result and the RL-controlled performance, as demonstrated in Figure \ref{extraction}. Therefore, the proposed EQRL algorithm exhibits superior training efficiency and convergence speed, making it highly suitable for real-time dynamic adjustments in AEF circuits. This also reflects that in fixed scenarios where the EMI spectrum remains fixed (without variations due to changes in operating conditions such as output power, temperature, or component ageing), although existing EMI suppression techniques can attenuate noise, the key advantage of our approach lies in its ability to self-tune and dynamically optimize the AEF parameters in response to variations in the noise spectrum. Conventional methods, which rely on fixed parameter settings or manual tuning, become ineffective when the EMI characteristics change, leading to suboptimal or degraded noise reduction performance over time.

\subsection{Evaluation of EQRL EMI Cancellation Effectiveness}
This section evaluates the offline performance of the trained policy when applied to untrained datasets, specifically Cable 44 and Cable 47. These datasets, characterized by distinct EMI noise profiles and broader frequency ranges, present an opportunity to assess the generalization and robustness of the proposed EQRL algorithm in managing unseen EMI conditions.
\subsubsection{Under untrained dataset}
\begin{figure}[h]
\centering
\includegraphics[width=0.485\textwidth]{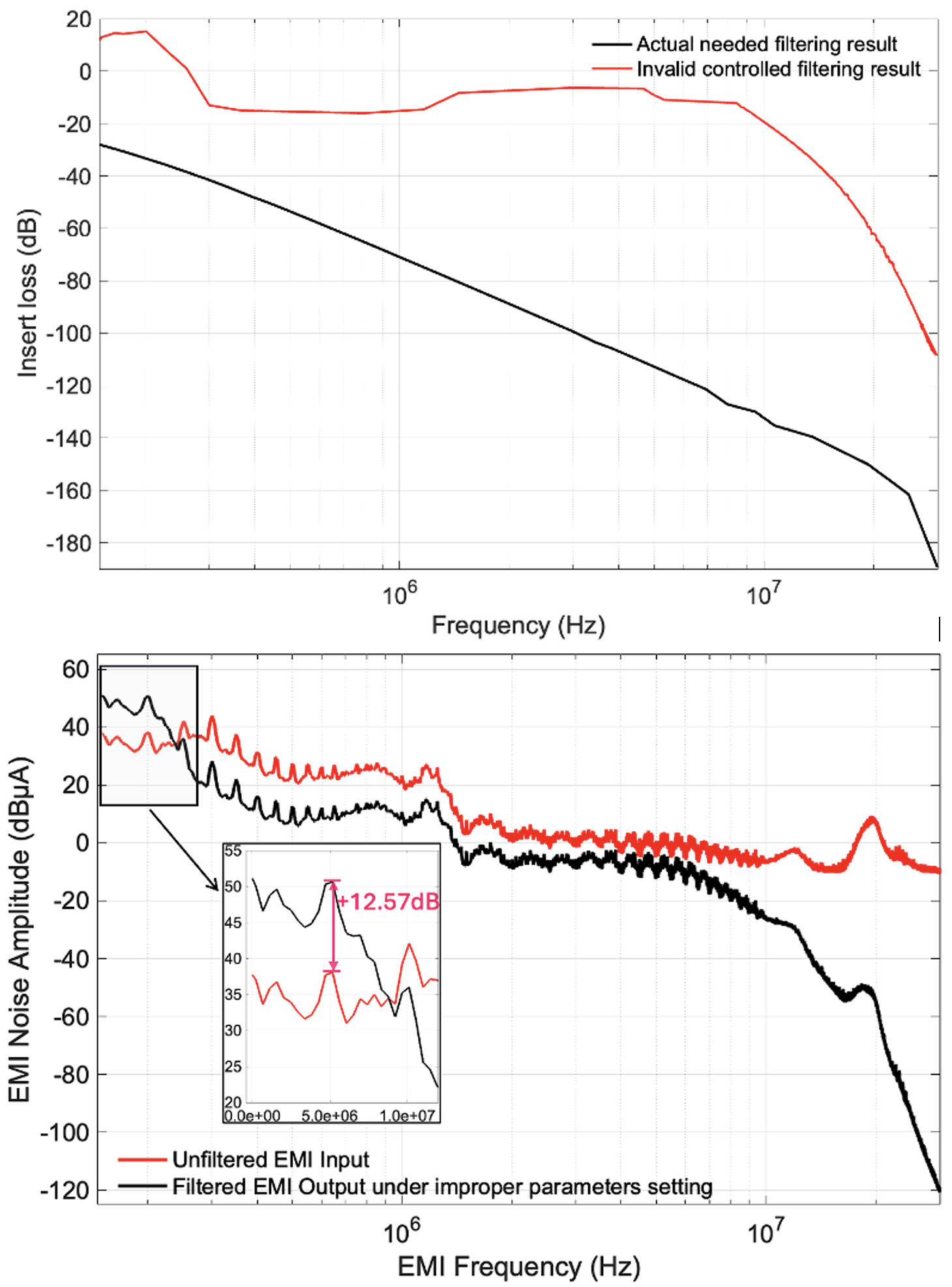}
\captionsetup{justification=centering} 
\caption{Cable 44 dataset insertion loss and EMI signal filtering performance without RL control}
\label{44_without}
\end{figure}

\begin{figure}[h]
\centering
\includegraphics[width=0.485\textwidth]{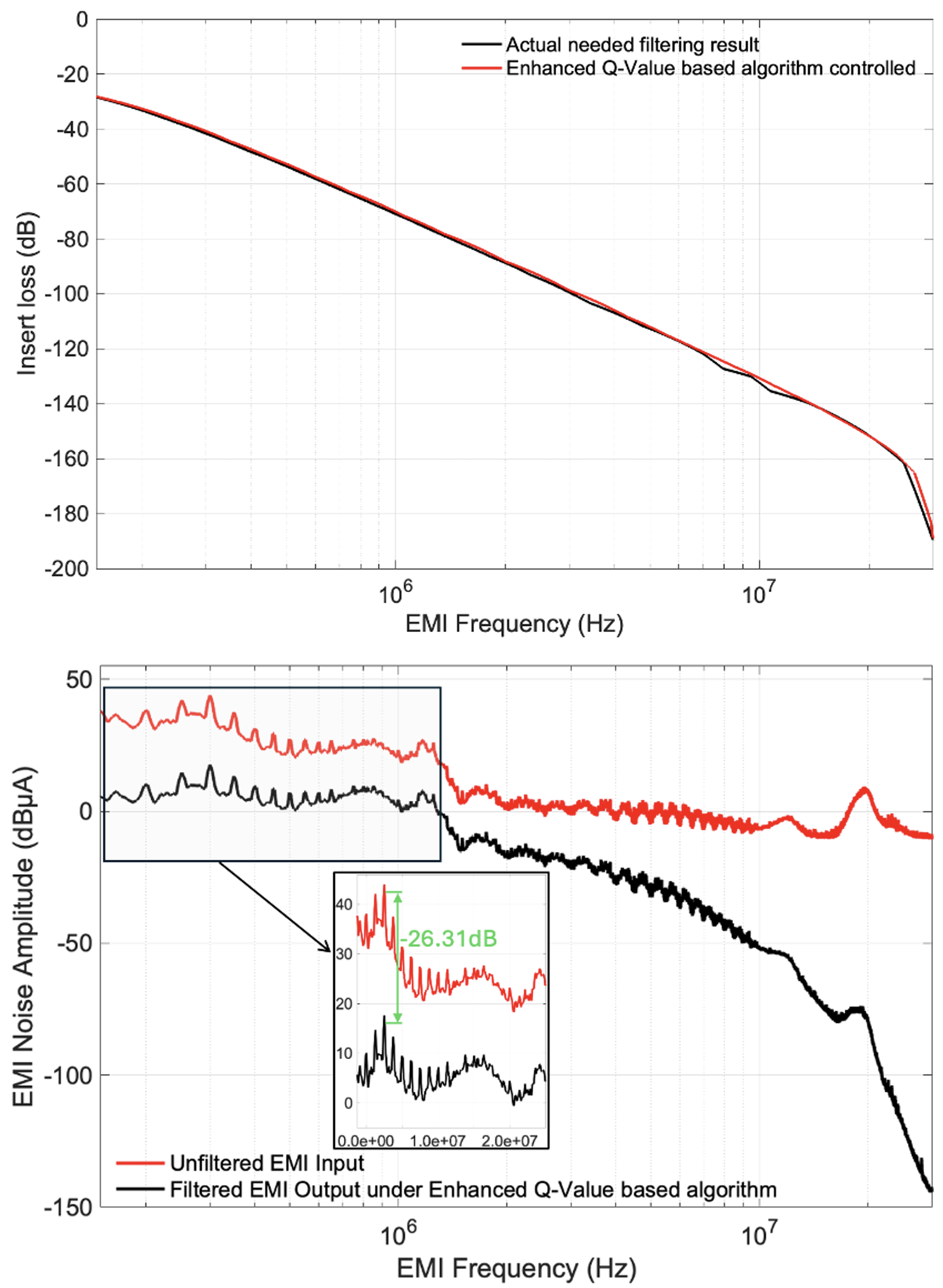}
\captionsetup{justification=centering} 
\caption{Cable 44 dataset insertion loss and EMI signal filtering performance}
\label{44_with}
\end{figure}

The Cable 44 dataset, covering a frequency range of 150 kHz to 30 MHz, simulates a typical automotive EMI scenario with varying noise levels. Figure \ref{44_without} illustrates the insertion loss performance without RL control, revealing significant fluctuations and insufficient EMI suppression across the bandwidth. However, when the RL-controlled policy was applied, as depicted in Figure \ref{44_with}, a notable improvement in EMI suppression was observed. The algorithm dynamically adjusted the AEF parameters, resulting in consistent filtering across the entire frequency range. As shown in Table \ref{Table 6}, the RMSE for the insertion loss under the EQRL algorithm on the Cable 44 dataset was 2.0502, with an average insertion loss of -29.15926 dB in the low-frequency range. This performance level demonstrates the algorithm’s ability to generalize effectively from the training dataset to the untrained Cable 44 dataset, ensuring robust filtering performance without requiring retraining or manual parameter adjustments. Figure \ref{47_with} also reflects that in fixed scenarios where the EMI spectrum remains fixed (without variations due to changes in operating conditions such as output power, temperature, or component ageing), although existing EMI suppression techniques can attenuate noise, the key advantage of our approach lies in its ability to self-tune and dynamically optimize the AEF parameters in response to variations in the noise spectrum. Conventional methods, which rely on fixed parameter settings or manual tuning, become ineffective when the EMI characteristics change, leading to suboptimal or degraded noise reduction performance over time.

\begin{table}[h!]
\centering
\caption{Filtering performance under enhanced Q-value based algorithm control}
\begin{tabular}{p{2cm} p{1.75cm} p{3.2cm}}
\toprule
\textbf{Datasets} & \textbf{Insertion loss RMSE} & \textbf{Average Insertion loss of low frequency part} \\
\midrule
Cable 44 & 2.05018792 & -29.15926 dB \\
Cable 47 & 2.89214039 & -26.47199 dB \\
Cable Bundle & 0.58991 & -30.88038 dB \\
\bottomrule
\end{tabular}
\label{Table 6}
\end{table}

\subsubsection{Under offline unknown scenarios}
\begin{figure}[h]
\centering
\includegraphics[width=0.485\textwidth]{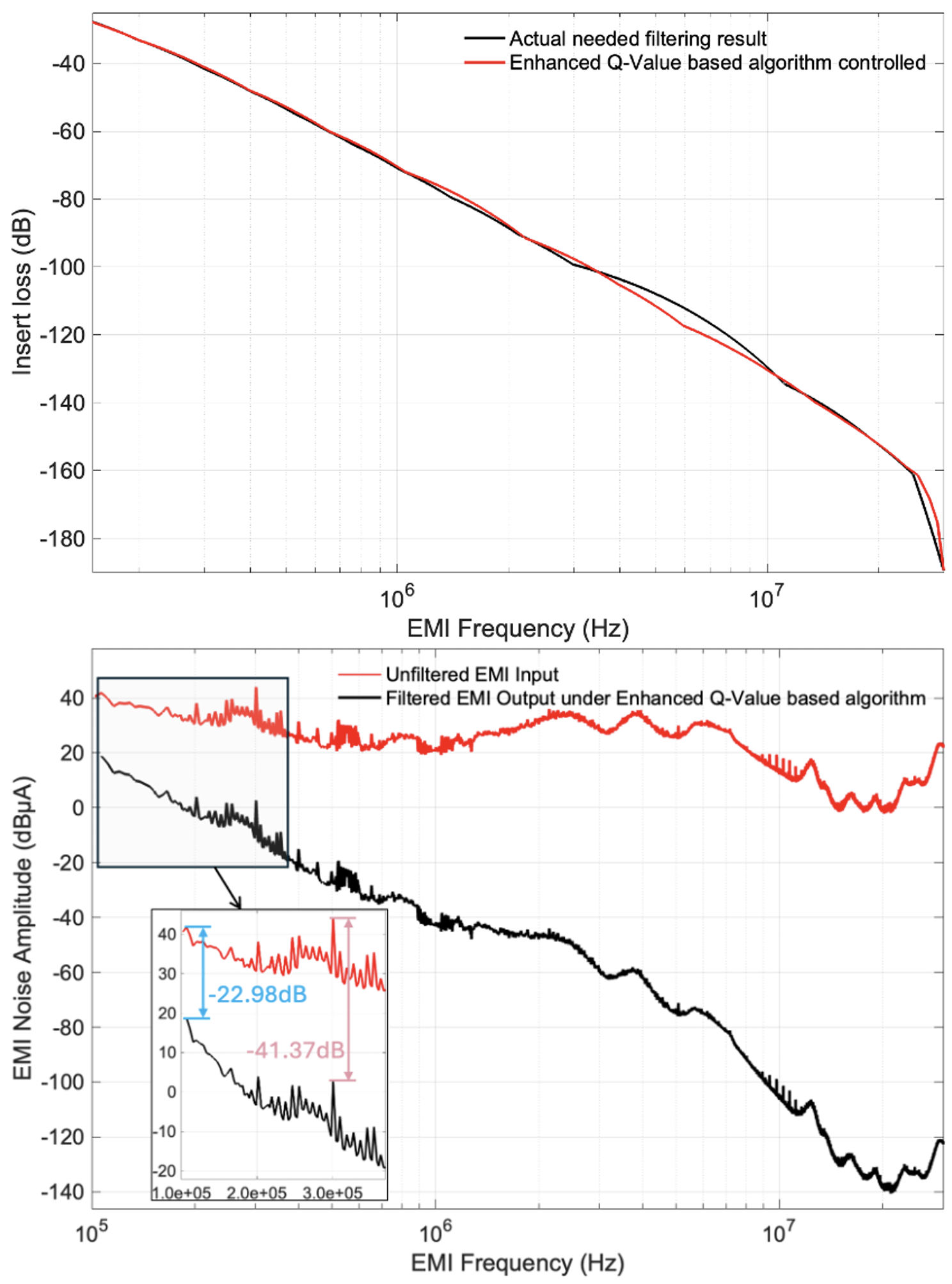}
\captionsetup{justification=centering} 
\caption{Cable 47 dataset with disturbance signal insertion loss and EMI signal filtering performance}
\label{47_with}
\end{figure}

The Cable 47 dataset introduces additional complexity due to more severe EMI interference and a broader noise spectrum ranging from 100 kHz to 30 MHz. This dataset encompasses lower frequency EMI compared to the Cable Bundle dataset used for training, which spans 150 kHz to 30 MHz. Additionally, interference signals between 100 kHz and 350 kHz were introduced to simulate uncertainties commonly encountered during vehicle operation. Figure \ref{47_with} demonstrates the performance of the EQRL-controlled AEF on this dataset, showcasing the algorithm’s capacity to adapt to higher noise levels and a broader frequency range. In comparison to the scenario without RL control, the RL-based approach achieved superior EMI suppression, particularly in filtering the lower frequency range, thus demonstrating improved generalization.

The RMSE for the Cable 47 dataset was 2.8921, slightly higher than that of Cable 44, but still within the acceptable range for effective EMI mitigation. The average insertion loss in the low-frequency range was -26.47199 dB, which, although marginally lower than that of Cable 44, still indicates robust suppression capability under challenging conditions.

\subsection{Trade-offs and Advantages}
To better illustrate the practical advantages of the proposed self-tuning active EMI filter (AEF), its performance and design characteristics are compared with those of commercially available passive EMI filters (PEFs). The representative PEFs follow standard LC and $\pi$-type topologies, which are the most common configurations used in automotive DC/DC converter applications \cite{Chu2016, Vedde2021}. These filters rely solely on fixed passive components and therefore provide effective but static attenuation performance within a limited frequency range. In contrast, the proposed reinforcement learning (RL)–controlled AEF integrates active compensation and adaptive parameter tuning, allowing the filter to dynamically adjust to changing EMI spectra and operating conditions. While classical PEFs typically achieve 10–15 dB attenuation in their designed frequency band, the proposed AEF consistently maintains 25–30 dB attenuation across a much wider frequency range (100 kHz–30 MHz). This improvement is particularly significant in dynamic EMI environments such as EDUs, where noise spectra shift due to variations in load, temperature, and switching conditions. However, the added complexity of sensing, processing, and control increases design and power requirements compared to passive filters. Therefore, the primary trade-off lies in system complexity and power consumption. The AEF requires auxiliary circuitry for sensing, digital control, and compensation, which introduces modest computational and energy overhead. However, advances in embedded hardware (e.g., DSPs and FPGAs) increasingly mitigate these costs, enabling practical real-time implementation in automotive platforms.

Table \ref{Table 7} summarises the key trade-offs and advantages of the proposed self-tuning active filter compared to commercially available passive EMI filters (PEFs). While commercially available PEFs, such as the TPSF12C1, offer simplicity, compactness, and cost-effectiveness, they rely on fixed parameters, which limit their adaptability to dynamic EMI conditions. While the current implementation focuses on a single tunable capacitor for simplicity and proof of concept, the proposed EQRL framework can be extended to accommodate multiple tunable components, such as feedback resistors, damping networks, or even gain settings across different AEF stages. In such multi-variable control scenarios, a multi-action RL formulation would be necessary, where the action space is defined over a vector of component adjustments rather than a scalar. The underlying EQRL structure, including the use of a VAE for latent state compression and a noise network for enhanced exploration, remains applicable; however, additional care must be taken to manage the increased dimensionality of both the state and action spaces. Techniques such as factorised action representations or actor-critic extensions may be integrated to maintain training stability and convergence efficiency.

\begin{table}[t!]
\centering
\caption{Key trade-offs and advantages of the proposed self-tuning active filter compared to commercially available passive EMI filters}
\begin{tabular}{>{\raggedright\arraybackslash}p{1.6cm} >{\raggedright\arraybackslash}p{2.75cm} >{\raggedright\arraybackslash}p{3.25cm}}
\toprule
\textbf{Feature} & \textbf{Commercially available PEFs} & \textbf{Proposed self-tuning AEFs} \\
\midrule
Performance & Fixed suppression, limited to the designed frequency range & Adaptive suppression, effective across varying EMI spectra \\
Flexibility & Static design, optimized for worst-case EMI & Dynamic adjustment based on real-time conditions \\
Adaptability & Limited, requires manual redesign for different applications & Self-tunes to different EMI sources and environments \\
Complexity & Simple, easy to implement & Requires additional sensing, control, and processing \\
Power consumption & Minimal (passive components) & Moderate due to active processing \\
Applicability & Suitable for general automotive EMI filtering & Ideal for advanced automotive power systems with dynamic EMI sources \\
\bottomrule
\end{tabular}
\label{Table 7}
\end{table}

Although the proposed EQRL-controlled AEF system has been validated through simulation, several practical considerations must be addressed for real-time hardware deployment. A primary challenge lies in achieving low-latency signal acquisition, decision-making, and actuation within the control loop. The RL agent, particularly when implemented with deep neural networks and VAE encoding, requires sufficient computational resources to operate within the tight time constraints of EMI-sensitive systems, typically on the order of microseconds to milliseconds. Embedded implementation would therefore necessitate efficient deployment on automotive-grade hardware such as Digital Signal Processors or Field-Programmable Gate Arrays with hardware-accelerated inference engines. Another critical bottleneck is the integration of high-precision, fast-response digital-to-analog interfaces for tuning passive components in real time, especially if using digitally controlled capacitors or voltage-variable components. Additionally, the cost and power overhead introduced by the sensing, computation, and actuation circuitry must be justified against the EMI suppression benefits. Despite these challenges, recent advances in edge AI and reconfigurable power electronics platforms suggest that real-time implementation of adaptive filtering strategies like EQRL is increasingly feasible, especially in high-end EV powertrains and autonomous systems where dynamic EMI conditions are prevalent. Therefore, future work will revolve around developing and experimentally evaluating a real-time prototype of the proposed self-tuning AEF on representative automotive platforms, such as on-board chargers (OBCs) and DC/DC converters, including full EMC testing and comparison against established passive and hybrid EMI filtering solutions to assess deployment feasibility in production environments.

\section{Conclusions}
This study introduces a novel approach for dynamically controlling active EMI filters (AEFs) in automotive systems using a proposed enhanced Q-value-based reinforcement learning (EQRL) algorithm. The research addresses the limitations of conventional passive and active filters in managing EMI in complex, dynamic environments. By developing a self-tuning AEF system, the proposed solution enables real-time adjustments to key circuit parameters, optimizing EMI suppression across a broad frequency range. 
The methodology employed in this study allows for more robust and adaptive EMI cancellation. Specifically, the use of deep RL, in conjunction with a variational autoencoder (VAE) for feature extraction, provided improved adaptability and generalization. This enabled the system to handle a wide variety of unpredictable EMI scenarios. Additionally, the integration of noise networks enhanced exploration efficiency, helping to avoid convergence on suboptimal solutions. Experimental results validated the effectiveness of the EQRL algorithm, demonstrating faster convergence rates compared to traditional RL algorithms such as SARSA and standard Q-learning. The algorithm’s ability to dynamically adjust key circuit components, such as the injection capacitor, resulted in superior filtering performance, as indicated by lower RMSE values and higher cumulative reward scores across both trained and untrained datasets.
The significance of this research lies in its practical application to automotive power systems, where dynamic EMI conditions are becoming increasingly common due to the rise of electrification and autonomous vehicle technologies. The proposed self-tuning AEF system not only enhances EMI suppression but also improves the overall reliability and stability of power systems in vehicles. By offering a scalable and adaptive solution, this work contributes to the broader field of EMI management and sets the stage for future advancements in intelligent EMI filtering technologies.  

\section*{Declaration of Competing Interest}
The authors declare that they have no known competing financial interests or personal relationships that could have appeared to influence the work reported in this paper.

\section*{Data availability}
Data will be made available on request.

\section*{Acknowledgments}
The authors would like to thank the financial support provided by the Royal Academy of Engineering through the Industrial Fellowship scheme (IF-2425-19-AI165).


\bibliographystyle{elsarticle-harv} 
\bibliography{cas-ref}

\end{document}